# ON THE QUEST FOR ECONOMIC PROSPERITY: A HIGHER EDUCATION STRATEGIC PERSPECTIVE FOR THE MENA REGION

Amr A. Adly

Electrical Power Engineering Dept., Faculty of Engineering, Cairo University, Giza 12613, Egypt

amradly@cu.edu.eg

adlyamr@gmail.com

http://orcid.org/0000-0001-5321-7877

**Abstract**

In a fast-changing technology-driven era, drafting an implementable strategic roadmap to achieve economic prosperity becomes a real challenge. Although the national and international strategic development plans may vary, they usually target the improvement of the quality of living standards through boosting the national GDP per capita and the creation of decent jobs. There is no doubt that human capacity building, through higher education, is vital to the availability of highly qualified workforce supporting the implementation of the aforementioned strategies. In other words, fulfillment of most strategic development plan goals becomes dependent on the drafting and implementation of successful higher education strategies. For MENA region countries, this is particularly crucial due to many specific challenges, some of which are different from those facing developed nations. More details on the MENA region higher education strategic planning challenges as well as the proposed higher education strategic requirements to support national economic prosperity and fulfill the 2030 UN SDGs are given in the paper.

## I. Introduction

There is no doubt that in a fast-changing technology-driven era, drafting an implementable strategic roadmap to achieve economic prosperity becomes a real challenge. While details and/or specifics of the national, regional or global strategic development plans may vary, they usually share some common goals. This can be easily realized by referring, for instance, to UN Sustainable Development Goals UN SDGs 2030, Africa Agenda 2063, and Egypt Vision 2030 (UN, n.d.), (Africa_Agenda_2063, 2015), (Egypt, n.d.). More specifically, ultimate goals usually include; improving citizens' quality of life and living standards (through boosting the national GDP per capita and its implications), and the creation of decent jobs.

It turns out that human capacity building, through higher education, is vital to the availability of highly qualified workforce supporting the implementation of the aforementioned strategies. Undoubtfully, quality higher education could play a crucial role in supporting the national economy in many ways (please refer to Figure 1). For instance, higher education institutions offer research and development (R&D) support to national industries that result in boosting the national value added, as previously demonstrated by (FISHER, 1939) and (Kapsos, 2006). Through proper curricula refinement, they could yield qualified graduates capable of implementing national development action plan as well as the acquisition of international job opportunities which would result in an influx of foreign currency remittances. Moreover, proper tuning of the different programs curricula could lead to building up of the graduates' entrepreneurial skills thus increasing the positive trend of private sector job creation. In addition, offering quality higher education programs, especially those which are internationally ranked and/or accredited, would result in the attraction of more international students thus leading to an influx of foreign currency. Finally, quality higher education would result in increasing the opportunities for national graduates to qualify for graduate scholarships in top international universities, the fact that could lead to an eventual inflow of highly qualified national talents or to the acquisition of more international job opportunities.

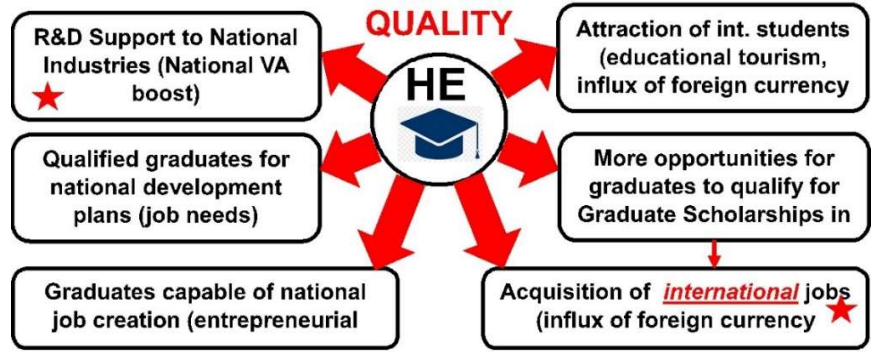

Figure 1. The role of quality higher education in supporting the national economy.

Obviously, fulfillment of most strategic development plan goals is dependent on the drafting and implementation of successful higher education strategies. This is especially crucial for MENA region countries due to many specific challenges, some of which are different from those facing developed nations. While the content of this article might be applicable to a number of developing nations, it presents a MENA region Higher Education strategic perspective. Obviously, drafting proper higher education strategies for the MENA region should take into consideration future global expectations as well regional and demographic specifics. In specific, higher education strategic planning for MENA region countries should take into consideration regional and international demographic expectations. Planning should also incorporate national and international job market needs in view of the fourth industrial revolution (4IR). These two issues stress the fact that no national strategic planning could be drafted in the absence of taking global factors into account. Hence, while drafting a strategy for higher education, recently compiled relevant reports by prominent international organizations such as; UNESCO, UNIDO, OECD, World Bank, WEF, McKinsey & Partners, …etc., should be explored. More details on the MENA region higher education strategic planning challenges as well as the proposed higher education strategic requirements to support national economic prosperity and fulfill the 2030 UN SDGs are given in the following sections.

**II. MENA Region Higher Education Strategic Planning Challenges**

In order to identify the salient features of successful higher education strategic planning for MENA region countries, main challenges would have to be pointed out. Those challenges are given below.

*Boosting the national value added*

Industries and/or business sectors are globally categorized as primary, secondary and tertiary (please refer to Figure 2). This classification is based on the value-added content and its consequent expected profit margin. Examples of primary industries include mining, primitive agriculture and mining. Secondary industries, sometimes referred to as blue collar industries, cover assembly industries and/or processing of primary industries. Examples include agro-food industries, liquification of natural gas, and petrochemical industries. It should be stated here that profit margin of secondary industries depend on the percentage of national components. Tertiary industries include service industries, whether low-tech or ICT-based. Examples include banking, retail, tourism and transportation. Such sectors are sometimes referred to as white collar industries. Tertiary industries also include high tech industries such as electronics and aerospace industries and are sometimes referred to as gold collar industries (FISHER, 1939), (Kapsos, 2006). Whether secondary or tertiary industries are considered, the percentage value-added is usually measured by the Competitive Industry Performance (CIP) index (UNIDO, 2018).

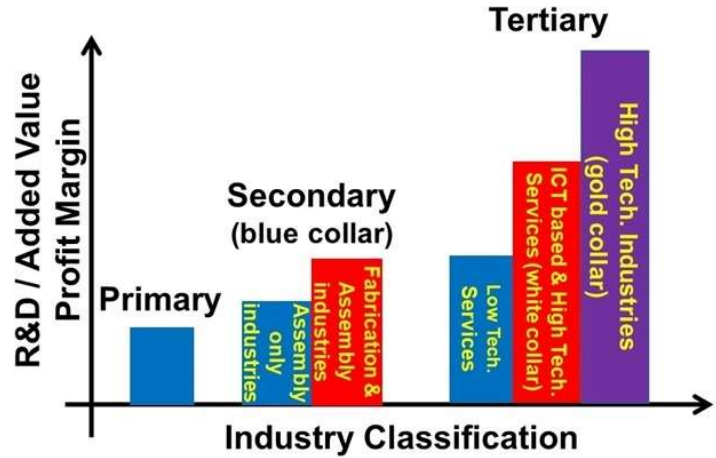

Figure 2. Global industry classification from the value-added/corresponding profit margin perspective.

In general, countries having high GDP per capita, high average salary per capita and high impact percentage of the overall global manufactured value-added share two main features. Specifically, these two features are a relatively high percentage of tertiary industry and a high CIP index. In other words, those countries posses high proportions of the so called knowledge-based economy" (Godin, 2006), (UNIDO, 2001). Currently, these two features are not common among MENA countries. Sample relevant data is shown in Figure 3 (WB, n.d.), (UNIDO, 2018), (UNIDO, 2019b), (UNIDO, 2020). This fact highlights the important role of building R&D capacities to support boosting national value-added components (UNIDO, 2019a). It is important to stress here that, unlike the situation in developed nations, most industry related R&D in the MENA region has to be outsourced through universities and research centers. This is clearly highlighted by the data shown in Figure 4 where the bold discrepancy between some MENA region countries and developed countries with regards to availability of full time equivalent (FTE) researchers as a percentage of the population as well as those affiliated to business enterprises (UNESCO Institute for Statistics, 2019). The above arguments highlight the increased expected role of higher education institutions in MENA region countries in comparison to the corresponding role in developed nations. This importance should be reflected in the specifics of any drafted strategy as will be discussed later.

*Global demographic expectations*

While drafting a higher education strategy, employability goals should be set straight. Targeting national job markets while ignoring international employability prospects might lead to the wrong design of curricula as well as negative economic consequences. A bold index that must be taken into consideration is the future expectations of the ageing index which represent the percentage ratio between population aged 65 and above to those aged below 15 (Frederick S. Pardee Center for International Futures, n.d.). In other words, such index gives a futuristic expectation of retirees to those about to enter the work force. Figure 5 demonstrates the expected ageing index values for some European and MENA region countries for the coming two decades. The same figure also presents the average ageing index expectations for the Europe as a whole and the MENA region countries. This figure suggests that while a workforce shortage is expected for most European countries, MENA region countries will face the challenge of massive job creation. This is especially evident from expected ageing index figure for Egypt where by 2040 the ageing index value will be about 33%. Alternatively, the corresponding value in Germany is expected to be approximately 250%. In addition to the global increasing life expectancy figures, these facts should be taken into consideration. More specifically, proper strategic planning for higher education in MENA region should include tailoring programs to prepare graduates for targeting as many international job openings as possible in addition to focusing on ageing related job needs such as; ageing medication, physical therapy, nursing and biomedical engineering (James Manyika et al., 2017).

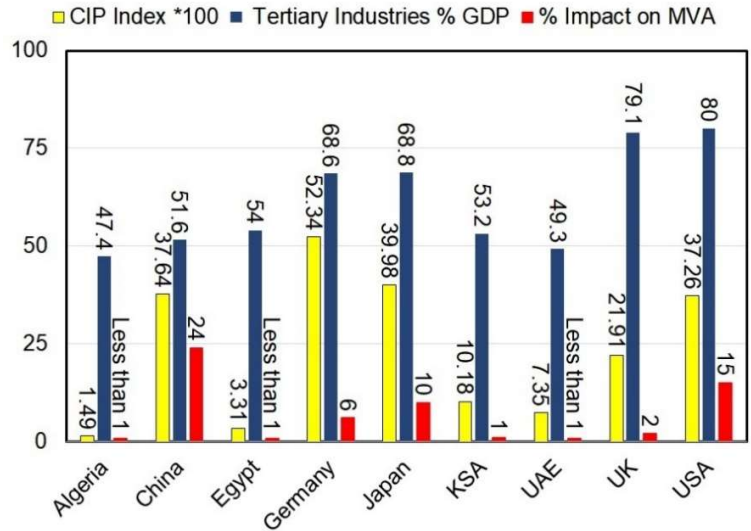

Figure 3. Data highlighting variations between some developed nations and some MENA countries with regards to the contribution of higher percentage of GDP generated by tertiary industries and CIP index on the impact on global manufactured value-added.

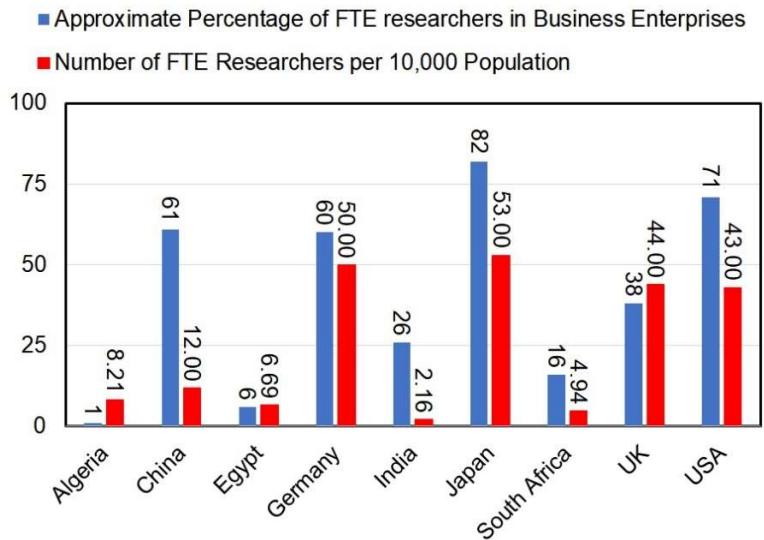

Figure 4. Data highlighting variations between some developed nations and some MENA countries with regards to the number of FTE researchers per 10,000 population as well as the percentage of FTE researchers in business enterprises.

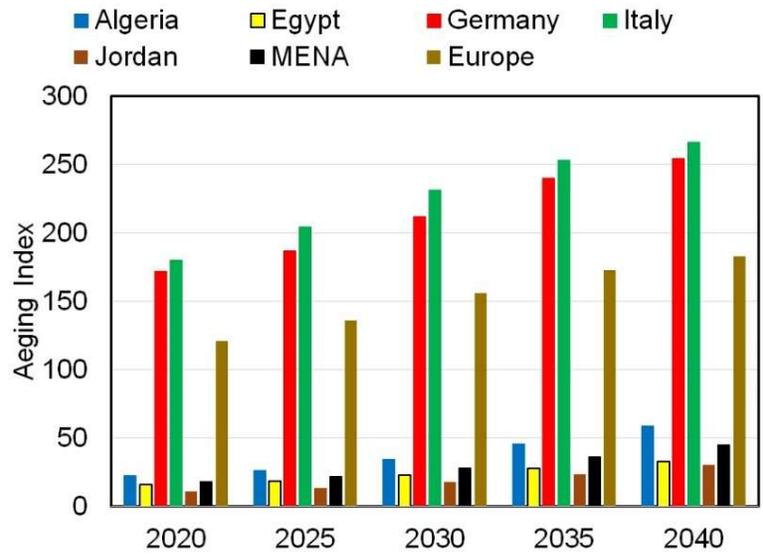

Figure 5. Aging index expectations for sample MENA and European countries as well as average ageing index expectations for European and MENA region countries.

*The fourth industrial revolution as a game changer of future jobs and their required qualifications*

      The fourth industrial revolution may be simply characterized by disruptive smart automation. Within this context, some jobs will be lost while other new jobs will emerge as shown in Figure 6 (James Manyika et al., 2017), (Bughin, Seong, et al., 2018), (Chui, M., Harryson, M., Mainyika, 2018), (Bughin, Hazan, et al., 2018), (Bughin et al., 2019), (Bryan Hancock, Kate Lazaroff-Puck, 2020). Facts suggest that the highest loss risk is concentrated on routine jobs with low skill (education) requirements and often low wages (James Manyika et al., 2017). Nevertheless, not all jobs are technically automatable especially those requiring social intelligence, non-routine jobs and those requiring high level of expertise. It is also expected that automation will mostly affect jobs related to industry, agriculture and some service sector jobs as shown in Figures 7 and 8. Unfortunately, entering the labor market for young people may become more difficult as entry level jobs have higher risk of automation.

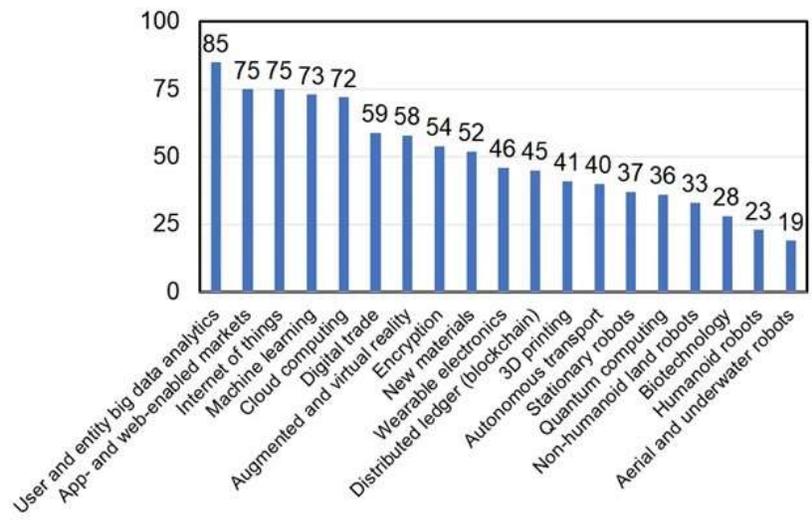

Figure 6. Estimated increase in adopted new technologies as a result of the fourth industrial revolution.

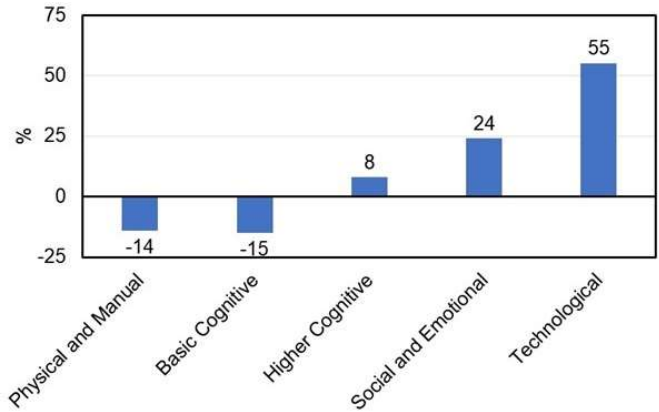

Figure 7. Estimated change in work hours spent over the period 2015-2030 as a result of the fourth industrial revolution.

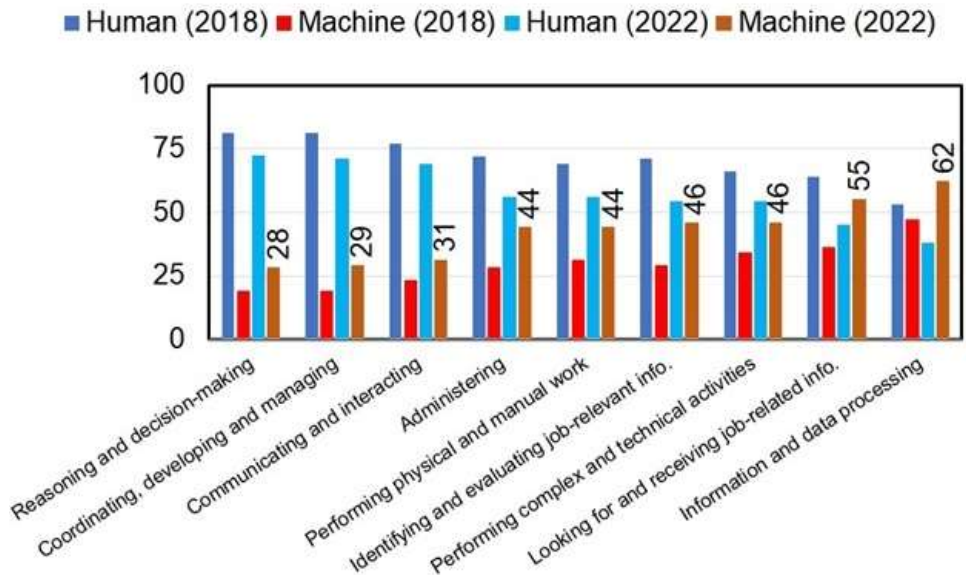

Figure 8. Expected evolution of human-machine job share as a result of the fourth industrial revolution.

It should also be stressed here that while overwhelming consequences of the fourth industrial revolution are not expected to affect developing countries before some years, indirect negative consequences on their national industries are expected as a result of the introduction of more competitive products and/or services (please refer to Figure 9). As previously stated, the fourth industrial revolution will also result in the creation of new jobs (Bughin, Seong, et al., 2018), (Manyika & Sneader, 2018), (OECD, 2018), (OECD, 2016), (Enders et al., 2019), (World Economic Forum, 2018), (OECD, 2017). Based upon studies related to future job needs in a number of developed nations, a higher demand is expected for jobs related to ageing medication, renewable energy, technology, computer and ICT. In fact, overall employment may end up on the rise provided that graduates have the right qualifications. Moreover, existing workforce displacement as a result of automation stresses the need for upskilling and/or reskilling to help current employees navigate the changing labor market needs. This further demonstrates the amplified importance of proper higher education strategic planning for MENA countries for both national and international employability prospects.

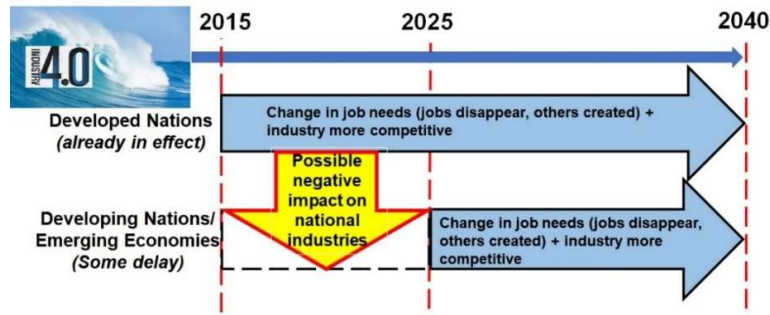

Figure 9. Expected high level implications of the fourth industrial revolution on developed and developing nations.

**III. Suggested MENA Region Higher Education Strategic Planning Features**

It has been clearly demonstrated that efforts directed to achieving the aspired economic prosperity in MENA region countries may be strongly supported through the proper higher education strategic planning (Schwab, 2019), (Wightman, 2020), (WEF, 2019), (World Economic Forum, 2020). Based upon the previously highlighted roles of higher education as well as the main challenges, a set of strategic planning features are suggested below.

*Immediate refinement of curricula of existing programs*

In view of the swift shifting future jobs skills requirements refinement of the curricula of existing programs is a must (UNESCO, 2015). This action should be applied on all education sectors, including Humanities and Social Sciences, and in accordance with expected both national and international future job prospects. This action should include the incorporation of basic skills related to UN SDG Goals across all higher education disciplines (Schleicher, 2018). (OECD, 2017). More specifically, those skills include:
- Foundation literacies (such as ICT, scientific, financial and cultural);
- Competencies (such as critical thinking, creativity, communication and collaboration);
- Character qualities (such as curiosity, initiative, persistence, adaptability and social as well as cultural awareness);

Curricula refinement should also include the incorporation of skills relevant to the fourth industrial revolution. Examples include; artificial intelligence (AI), big data, blockchain and internet of things (IoT) (OECD, 2016), (Manyika & Sneader, 2018), (Enders et al., 2019). This curricula refinement is imperative even for social sciences and humanities disciplines since graduates of these disciplines represent the majority of many service industries workforce. Moreover, given the fact that preparing graduates to target international job openings should be on MENA region strategic goals, it might be necessary to align higher education systems in MENA countries with analogous international systems such as the EU Bologna Higher Education System.

*Initiation of non-existent programs that fulfil fourth industrial revolution futuristic jobs*

There is no doubt that totally new jobs will emerge as a result of the fourth industrial revolution (OECD, 2016), (World Economic Forum, 2017), (Jahanian, 2020), (Schleicher, 2018), (Nordlund, 2020), (Enders et al., 2019), (Chui, M., Harryson, M., Mainyika, 2018). Like for the case of many countries worldwide, higher education programs serving the qualification needs of such futuristic jobs would have to be initiated in MENA countries. In other words, strategic planning of higher education programs should take this into consideration. Obviously, many of those programs have already been initiated in top ranked universities situated in developed nations. While the major content components of some of those programs are novel, many of such programs are basically interdisciplinary in nature. Sample examples of those programs include:

- Brain and cognitive sciences;
- Computational and systems biology;
- Agricultural and consumer economics;
- Mathematical economics;
- Computer science economics and data science;
- Humanities and engineering and sciences;
- Strategy and innovation management;
- Quantum computing;
- Technology, operations and value chain management.

*Incorporation of research and development capacity building curricula*

In addition to the need for boosting the research facilities and infra structures, incorporation of research and development capacity building curricula is crucial to support national efforts aiming at boosting the value added in the various national industrial and service sectors (UNIDO, 2019a), (WB, n.d.). As previously pointed out in challenges related to the implications of the fourth industrial revolution, this is also imperative to support the competitiveness edge of national products with regards to imported products. Curricula-wise, this includes resorting to novel teaching strategies such; research-based education, competition-based education and industry-mandated design projects. The general ultimate goal is to offer expertise necessary to increase the percentage GDP contribution of tertiary industries characterized by high profitability margins. Obviously, every MENA nation is expected to focus on specific specialties relevant to its existing and planned production sectors. A possible road map of actions required to be taken into account during the strategic planning of higher education programs to boost national value added in production sectors is shown in Figure 10 (UNIDO, 2019a), (UNIDO, 2020), (Schwartz, 2019). It should be mentioned that international publications ranking of some MENA countries reflect the availability of high caliber researchers in local universities and research centers. Examples include Kingdom of Saudi Arabia and Egypt which were internationally ranked for the 2018 publications in the 32nd and 35$^{th}$ position, respectively (Scimago, 2019).

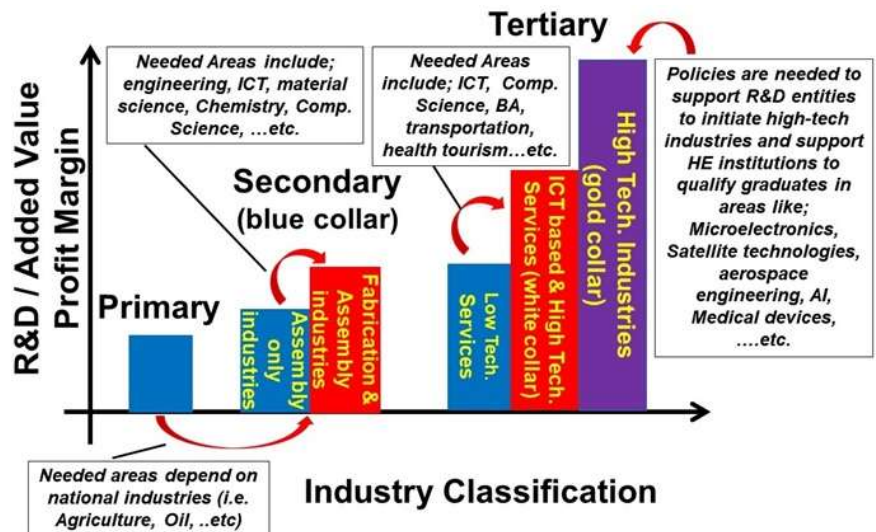

Figure 10. A possible road map of actions required to be taken into account during the strategic planning of higher education programs to boost national value added in production sectors.

*Supporting actions aiming at boosting International Ranking and Accreditation*

Having highlighted the need for empowering MENA graduates to acquire more international jobs (Frederick S. Pardee Center for International Futures, n.d.), institutional international ranking improvement and international accreditation of programs turn to be important from national economy perspectives as previously discussed as well as demonstrated by Figure 11. According to the recently publicized QS World University Ranking, only about 40 MENA universities are listed among the top ranked 1000 world universities (QS, n.d.). Hence, more actions would have to be taken along this front. It should also be mentioned that international accreditation of programs could be complemented by joint and/or double degree programs with prominent international institutes

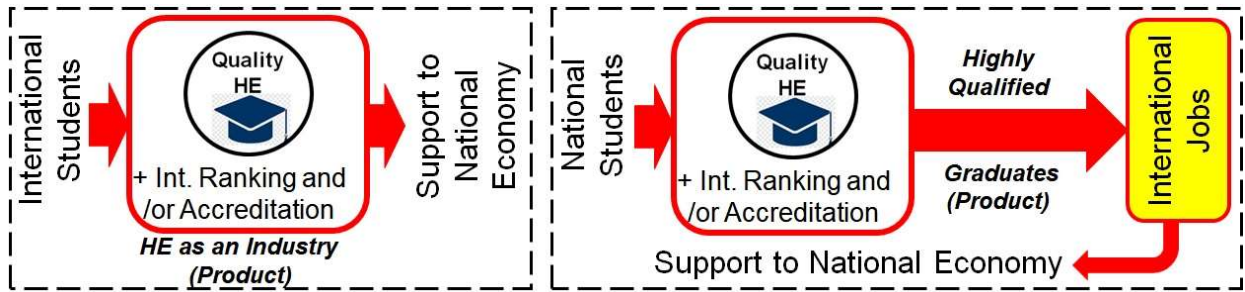

Figure 11. International ranking and accreditation as supporting components to the national economy.

## IV. Concluding Remarks

It can be concluded from the previous analysis and discussions that appropriate higher education strategic planning in MENA region countries is crucial for supporting the quest to achieve prosperous economies. In other words, allocating funds for appropriately planned higher education sectors in MENA region countries should not be considered as expenditure but rather as investments with high return expectations. Strategic planning for higher education in MENA region countries should also involve the following aspects:

- Taking into consideration both national and international future job needs, the fact that necessitates adaptation of higher education strategies in accordance with the fourth industrial revolution requirements;
- Initiation of non-existent programs relevant to new future fourth industrial revolution jobs;
- Adapting curricula of existing programs to future national and international job requirements in addition to the incorporation of relevant UN SDG skills;
- Alignment with international higher education systems and frameworks;
- Boosting international ranking and accreditation from a national economy support perspective.

Moreover, it should be mentioned that taking the economic perspective while considering higher education strategic planning leads to more harmonization with UN SDGs as shown in the Table I. Finally, while the content of this paper addresses strategic higher education planning for MENA countries, it is believed that considerable components of this perspective is applicable to a large number of nations classified as developing countries and/or emerging economies.

Table I. Correlation between aspects related to economic perspective of strategic higher education planning and UN SDGs

| Strategic Higher Education Planning Aspects | Relevant UN SDGs (UN, n.d.) |
|---|---|
| Targeting quality up-to-date higher education programs | GOAL 4: Quality Education |
| Rightfully realizing the return on investment in higher education on the national economy should justify more national budget allocation to increase fair access opportunities to quality higher education programs | GOAL 1: No Poverty, GOAL 5: Gender Equality, GOAL 8: Decent Work and Economic Growth, and GOAL 10: Reduced Inequality |
| Adopting the aforementioned higher education strategies should result in programs that fully support national economies across a very wide spectrum of goals including boosting national added value, initiation of high-tech and ICT-based service industries in addition to responding to national and environmental challenges | GOAL 2: Zero Hunger, GOAL 3: Good Health and Well-being, GOAL 6: Clean Water and Sanitation, GOAL 7: Affordable and Clean Energy, GOAL 9: Industry, Innovation and Infrastructure, GOAL 11: Sustainable Cities and Communities, GOAL 13: Climate Action, GOAL 14: Life Below Water, and GOAL 15: Life on Land |